# Vernier-assisted mode-selective PT symmetry in optoelectronic oscillators

Gazi M. Hasan[1], Abhijit Banerjee[2] & Trevor J. Hall[1]

[1]University of Ottawa, Electronic Engineering & Computer Science, Advanced Research Complex, 25 Templeton St, Ottawa, Ontario, K1N 6N5, Canada

[2]Academy of Technology, Department of Electronics & Communications Engineering, Adisaptagram, Hooghly, West Bengal, 712121, India.

## Abstract

Mode-selective $\mathcal{PT}$-symmetry is the manifestation of Vernier-assisted cross-injection between oscillators with unequal delays. $\mathcal{PT}$-symmetry has been proposed as a mechanism for achieving low-noise single-mode operation in optoelectronic oscillators, but existing formulations are typically based on matched delay loops and frequency-independent coupling, leading to symmetry transitions that are global in frequency and therefore not intrinsically mode selective. A more general formulation of $\mathcal{PT}$-symmetric time-delay oscillators is developed in which the coupling operator is allowed to be dispersive. This permits frequency-selective $\mathcal{PT}$-symmetry transitions and removes the requirement for matched delay loops. Two mathematically equivalent but physically distinct realisations are derived. The first corresponds to coupled gain-loss loops connected by a dispersive coupler, while the second corresponds to a pair of equal-gain oscillators coupled through symmetric cross-injection with unequal delays. Numerical simulations confirm the predicted behaviour and demonstrate strong sidemode suppression while preserving the low phase-noise characteristics of large-delay oscillators. The results establish a direct connection between $\mathcal{PT}$-symmetry, cross-injection architectures, and Vernier oscillator design.

## 1. Introduction

Accurate control of time and frequency is central to advances in optical and wireless communications, as well as in radar, lidar, aerospace systems, and precision metrology. The performance of these applications is ultimately limited by noise and instability in their underlying oscillators. Time-delay oscillators (TDOs), including lasers [1] and optoelectronic oscillators (OEOs) [2], exploit delayed feedback to suppress noise and enhance spectral purity. In such systems, long delays—realised through extended optical cavities or fibre-optic delay lines—lead to dense mode spectra and strongly influence both stability and phase-noise characteristics.

Parity–time ($\mathcal{PT}$) symmetry has been explored in optical and laser systems [3-5] and, more recently, in optoelectronic oscillators [6,7], with the aim of achieving robust single-mode operation while maintaining ultra-low phase noise and strong sidemode suppression. Discrepancies between theoretical predictions and experimental observations have motivated the development of delay-differential-equation and delay-difference formulations [8-13], together with numerical tools for modeling injection-locked and phase-locked TDOs, including $\mathcal{PT}$-symmetric configurations.

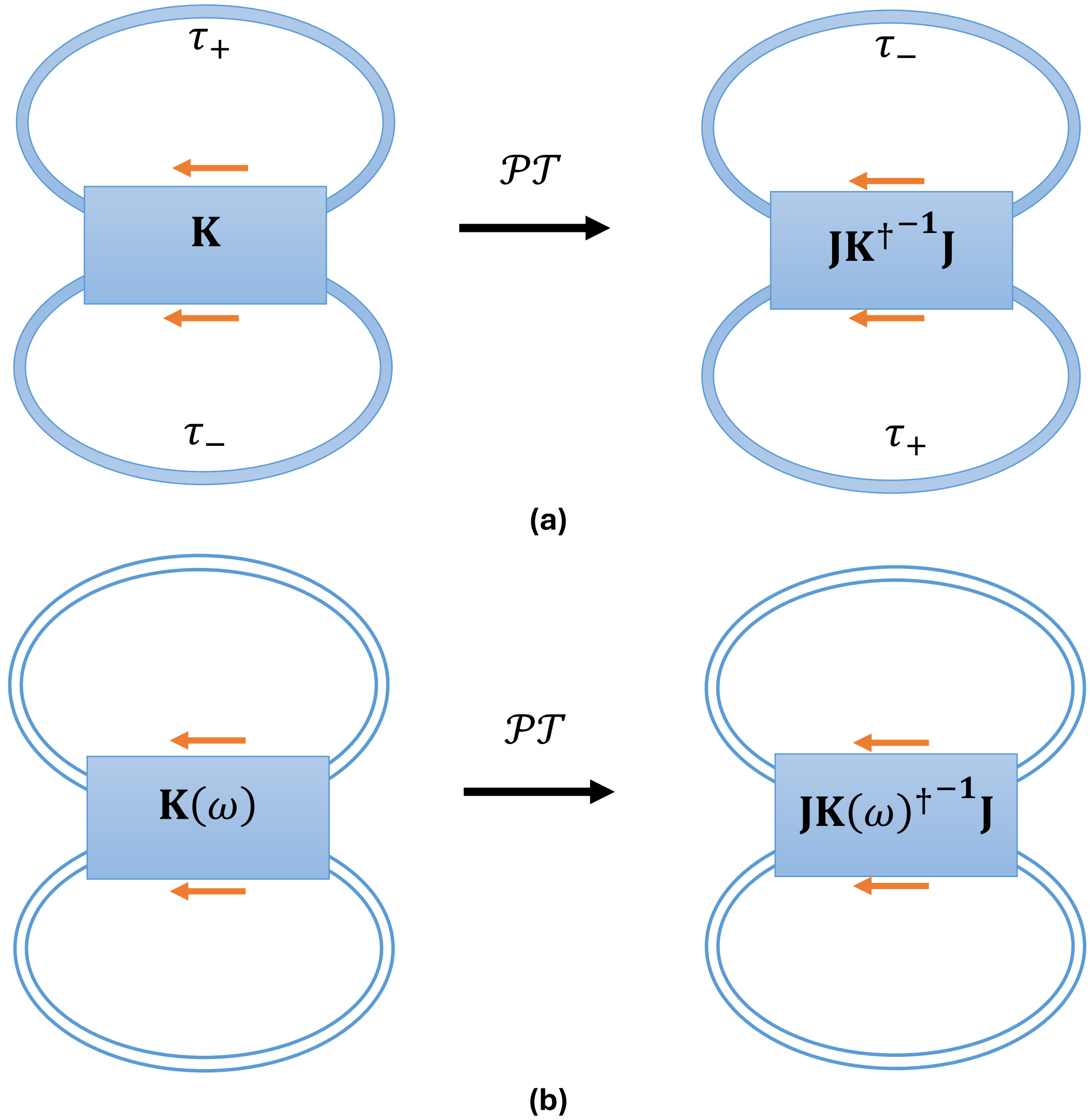

*Figure 1 $\mathcal{PT}$ symmetry applied to a time delay oscillator modeled by: (a) a pair of delay loops interconnected by a coupler with transmission matrix* $\mathbf{K}$, *to (b) a round-trip operator with dispersive transmission* $\mathbf{K}(\omega)$ *with zero-delay feedback connections. Model (a) naturally suggests the interpretation that PT symmetry requires identical delays because the two loops are interchanged by the parity-time operator. However, the transmission matrix of a pair of delay loops with different delays satisfies $\mathcal{PT}$-symmetry. The key insight is that a dispersive coupler is necessary to describe the differential component of the delay. Once one allows the coupler to be dispersive, consideration of the common part of the delays is redundant, yielding model (b).*

In earlier work [13], a non-perturbative framework for $\mathcal{PT}$-symmetric time-delay oscillators was developed based on a delay–difference description combined with a scattering-matrix representation of the coupling network. Within this approach, the modal structure of the system is determined in closed form without reliance on perturbative approximations. Under the simplifying assumption of unidirectional propagation, the full scattering problem reduces to an effective two-dimensional transmission model, leading to an interpretation in which the $\mathcal{PT}$-symmetry phase transition appears as a global property of the oscillator and is not mode-selective.

$\mathcal{PT}$-symmetry does not in general require that individual subsystems independently satisfy the symmetry condition. A broader formulation is obtained by allowing the coupling operator to include dispersion, so that it represents a frequency-dependent round-trip transformation. In the Fourier domain this corresponds to a dispersive transmission matrix, which is formally equivalent to replacing matched delay loops with zero-delay feedback paths. This conceptual shift is illustrated in Figure 1.

Within this extended framework, imbalanced effective delays arise naturally, and localised, mode-selective $\mathcal{PT}$-symmetry phase transitions become possible. Mode-selective $\mathcal{PT}$-symmetry is the manifestation of Vernier-assisted cross-injection between oscillators with unequal delays. This interpretation connects $\mathcal{PT}$-symmetric dynamics directly to practical dual-loop oscillator architectures, bridging formal symmetry concepts and device-level design.

The present work makes two principal contributions. First, it is shown that mode-selective $\mathcal{PT}$-symmetry transitions require dispersive coupling, and that $\mathcal{PT}$-symmetry does not require matched delay loops. Second, two physically distinct but mathematically equivalent representations of mode-selective $\mathcal{PT}$-symmetric oscillators are developed. One corresponds to coupled gain-loss loops with dispersive coupling, while the second corresponds to cross-injection between equal-gain oscillators with unequal delays. The latter admits a direct Vernier interpretation. Theoretical predictions are validated using the authors' custom delay-differential-equation simulation platform.

The remainder of the paper is organised as follows. Section 2 applies $\mathcal{PT}$-symmetry to the roundtrip operator $\mathbf{K}$ illustrated by two realisations $\mathbf{K_I}$, $\mathbf{K_{II}}$ that support frequency dependent $\mathcal{PT}$-symmetry transitions and which are shown to be equivalent while differing in interpretation. Section 3 discusses the implementation architecture of the two realisations: $\mathbf{K_I}$ as a dispersive-coupled gain-loss matched loop system, and $\mathbf{K_{II}}$ as a cross-injection mismatched delay loop system. Section 4 presents simulation results obtained using a Simulink model of the cross-injection architecture. Section 5 summarises the key findings, discusses their implications and draws conclusions. Appendix I provides mathematical details supporting the main text.

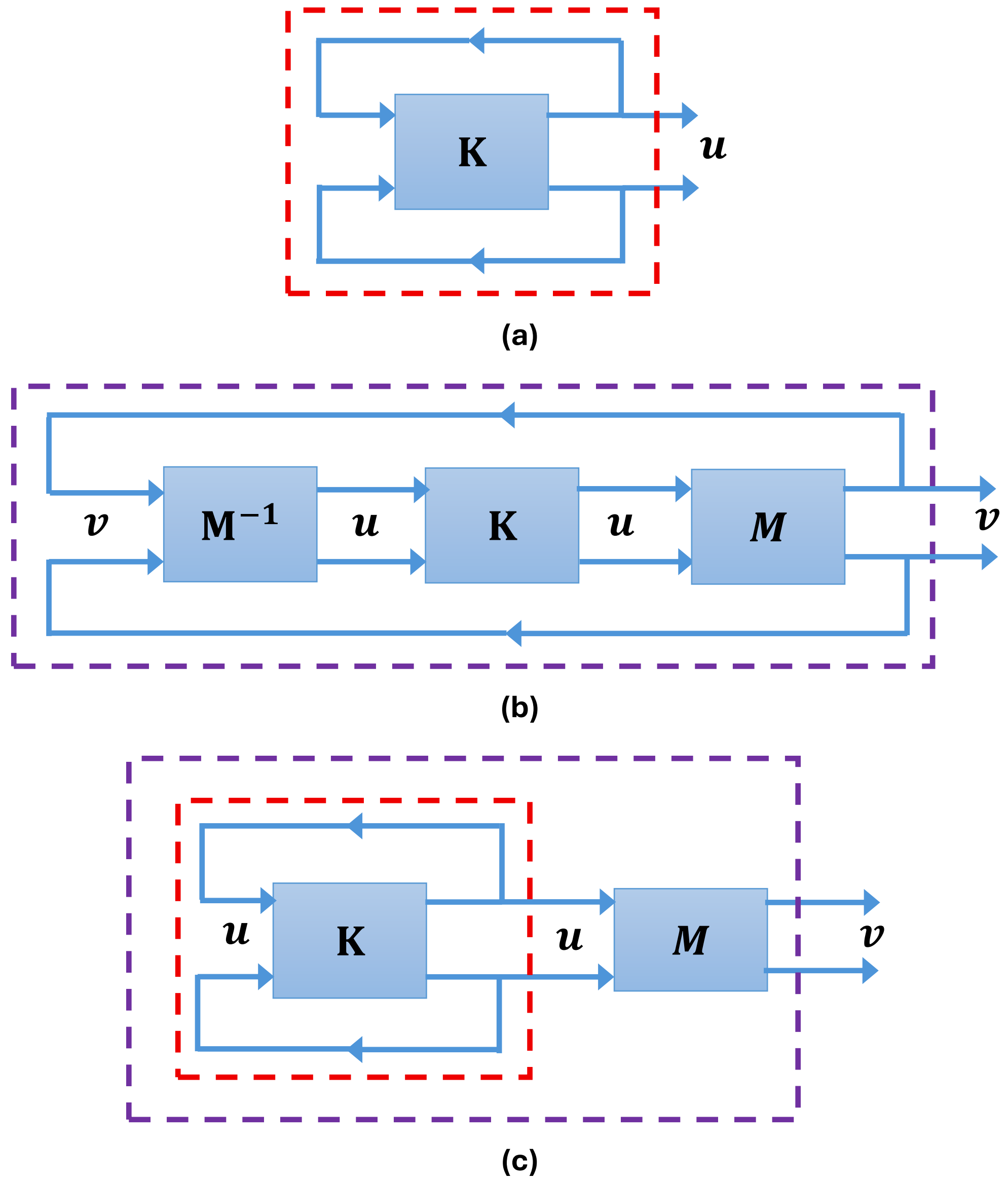

*Figure 2 Illustration of transformation of a roundtrip operator by similarity demonstrating preservation the core oscillator structure. (a) An oscillator system enclosed in a red-dashed box described by the roundtrip operator* $\mathbf{K}$*. (b) An oscillator system enclosed in a purple-dashed box described by the similarity transformation* $\mathbf{MKM}^{-1}$ *of the roundtrip operator* $\mathbf{K}$ *where* $\mathbf{M}$ *is an invertible matrix but otherwise arbitrary. (c) A physically equivalent oscillator system to the transformed oscillator simplified by the cancellation of* $\mathbf{M}$ *and* $\mathbf{M}^{-1}$*within the feedback loop, which restores the original oscillator shown enclosed by the red-dashed box. The internal oscillator state vector* $\boldsymbol{u}$ *may be observed externally through the transformed state vector* $\boldsymbol{v}$*.*

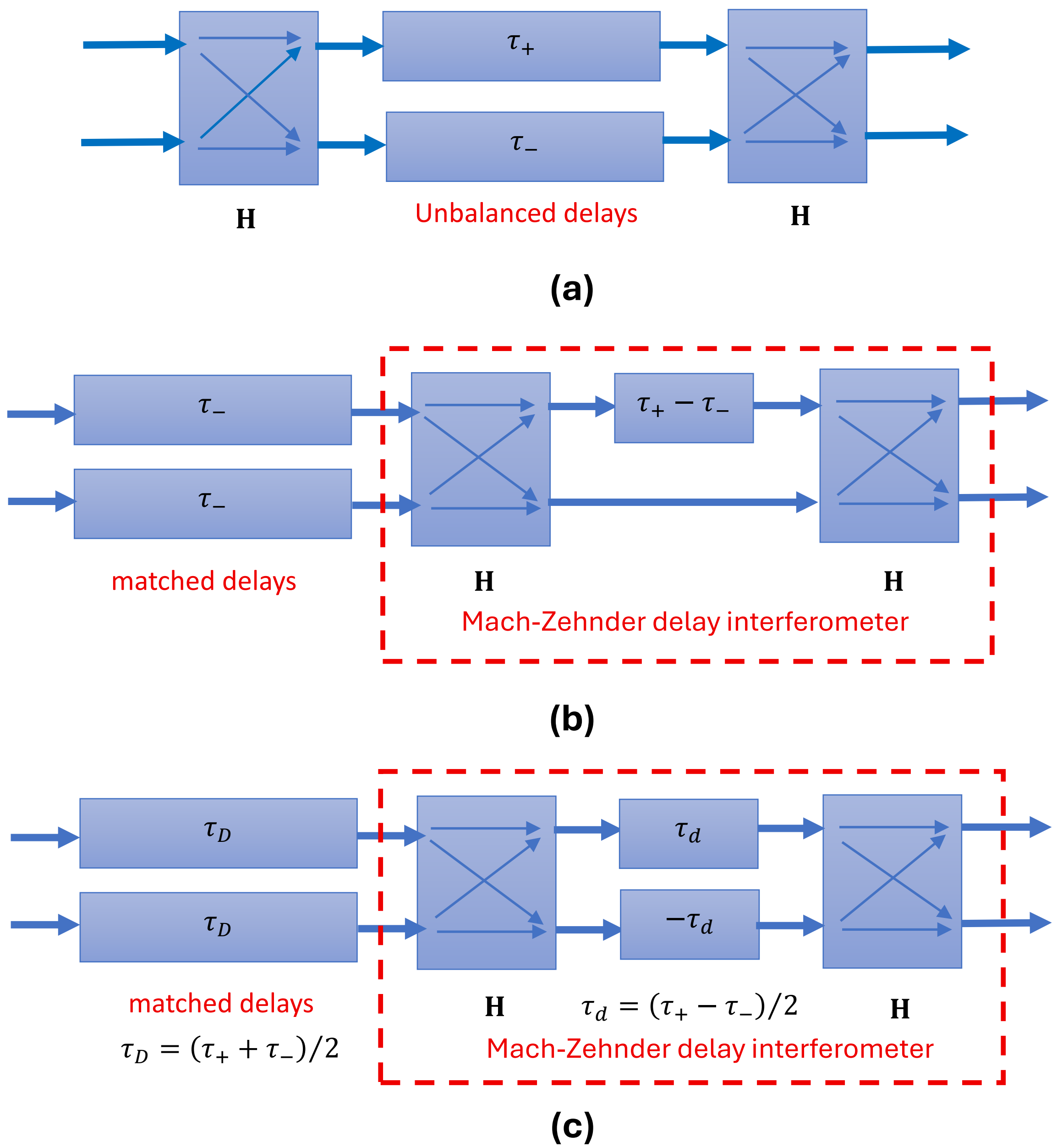

*Figure 3 An illustration of the equivalence of (a) an unbalanced pair of delay lines with delay times $\tau_+ > \tau_-$ placed between two $2 \times 2$ Hadamard couplers $\boldsymbol{H}$ to (b) & (c) a balanced pair of delay lines concatenated with a Mach-Zehnder delay interferometer. A physical implementation must use causal components as in (b). For analysis, one is free to use noncausal components provided the complete subsystem is causal as in (c). The Mach-Zehnder delay interferometer in (c) acts as a dispersive coupler. It has a transmission matrix of the same form as a lossless variable coupler $\boldsymbol{U}(\theta)$ but with a splitting angle $\theta(\omega) = -\omega\tau_d$ linear in frequency $\omega$.*

## 2. Dispersive $\mathcal{PT}$-symmetric oscillator model

A previous work [13] showed that a scattering matrix $\mathbf{S}$ is $\mathcal{PT}$-symmetric if:

$$\mathbf{SJS}^{\dagger} = \mathbf{J}$$

*Equation 1*

where $\mathbf{J}$ is the exchange matrix that has a trailing diagonal with unit elements only and all zeros elsewhere. Under the assumption of unidirectional propagation this condition applies to a transmission matrix $\mathbf{K}$ with one-half the dimension of the scattering matrix. In the simplest case of two dimensions $\mathbf{J} = \sigma_1$, where $\sigma_1$ is the first Pauli matrix. Consequently, it is useful to introduce the Pauli matrices:

$$\sigma_1 = \begin{bmatrix} 0 & 1 \\ 1 & 0 \end{bmatrix} \quad ; \quad \sigma_2 = \begin{bmatrix} 0 & -i \\ i & 0 \end{bmatrix} \quad ; \quad \sigma_3 = \begin{bmatrix} 1 & 0 \\ 0 & -1 \end{bmatrix}$$

*Equation 2*

and their linear combination into the Hadamard matrix:

$$\mathbf{H} = \frac{1}{\sqrt{2}}(\sigma_1 + \sigma_3) = \frac{1}{\sqrt{2}}\begin{bmatrix} 1 & 1 \\ 1 & -1 \end{bmatrix}$$

*Equation 3*

The Hadamard matrix is a real symmetric unitary and therefore *involutory* matrix which plays a major role in the analysis as it diagonalizes the exchange matrix:

$$\mathbf{HJH} = \sigma_3$$

*Equation 4*

In Appendix I two representations $\mathbf{K_I}$, $\mathbf{K_{II}}$ of a $\mathcal{PT}$-symmetric roundtrip operator $\mathbf{K}$ are constructed:

$$\mathbf{K_I}(\omega) = \mathbf{\Gamma}(\gamma)\mathbf{U}(\theta_0 - \omega\tau_d)\mathbf{D}_{\tau_D,\tau_D}(\omega)$$

*Equation 5*

$$\mathbf{K_{II}}(\omega) = \mathbf{\Sigma}(\gamma)\mathbf{\Theta}(\theta_0)\mathbf{D}_{\tau_+,\tau_-}(\omega)$$

*Equation 6*

where the component matrix factors are defined by:

$$\mathbf{D}_{\tau_+,\tau_-}(\omega) = \begin{bmatrix} \exp(-i\omega\tau_+) & 0 \\ 0 & \exp(-i\omega\tau_-) \end{bmatrix} = \exp(i\omega\tau_d\sigma_3)\,\mathbf{D}_{\tau_D,\tau_D}(\omega)$$

*Equation 7*

$$\mathbf{\Theta}(\theta) = \exp(i\theta\sigma_3)$$

*Equation 8*

$$\mathbf{U}(\theta) = \mathbf{H\Theta}(\theta_0)\mathbf{H}$$

*Equation 9*

$$\mathbf{\Gamma}(\gamma) = \exp(\gamma\sigma_3)$$

*Equation 10*

$$\mathbf{\Sigma}(\gamma) = \mathbf{H}\mathbf{\Gamma}(\gamma)\mathbf{H} = \begin{bmatrix} \cosh(\gamma) & \sinh(\gamma) \\ \sinh(\gamma) & \cosh(\gamma) \end{bmatrix}$$

*Equation 11*

The two representations $\mathbf{K_I}$, and $\mathbf{K_{II}}$ are simply related by a similarity relation:

$$\mathbf{K_I} = \mathbf{H}\mathbf{K_{II}}\mathbf{H}$$

*Equation 12*

which preserves the eigenvalue structure. Moreover, in the case of an oscillator roundtrip operator, it also conserves the *physical structure* of the core oscillator, only the external observation of the oscillator state vector is transformed as illustrated in Figure 2.

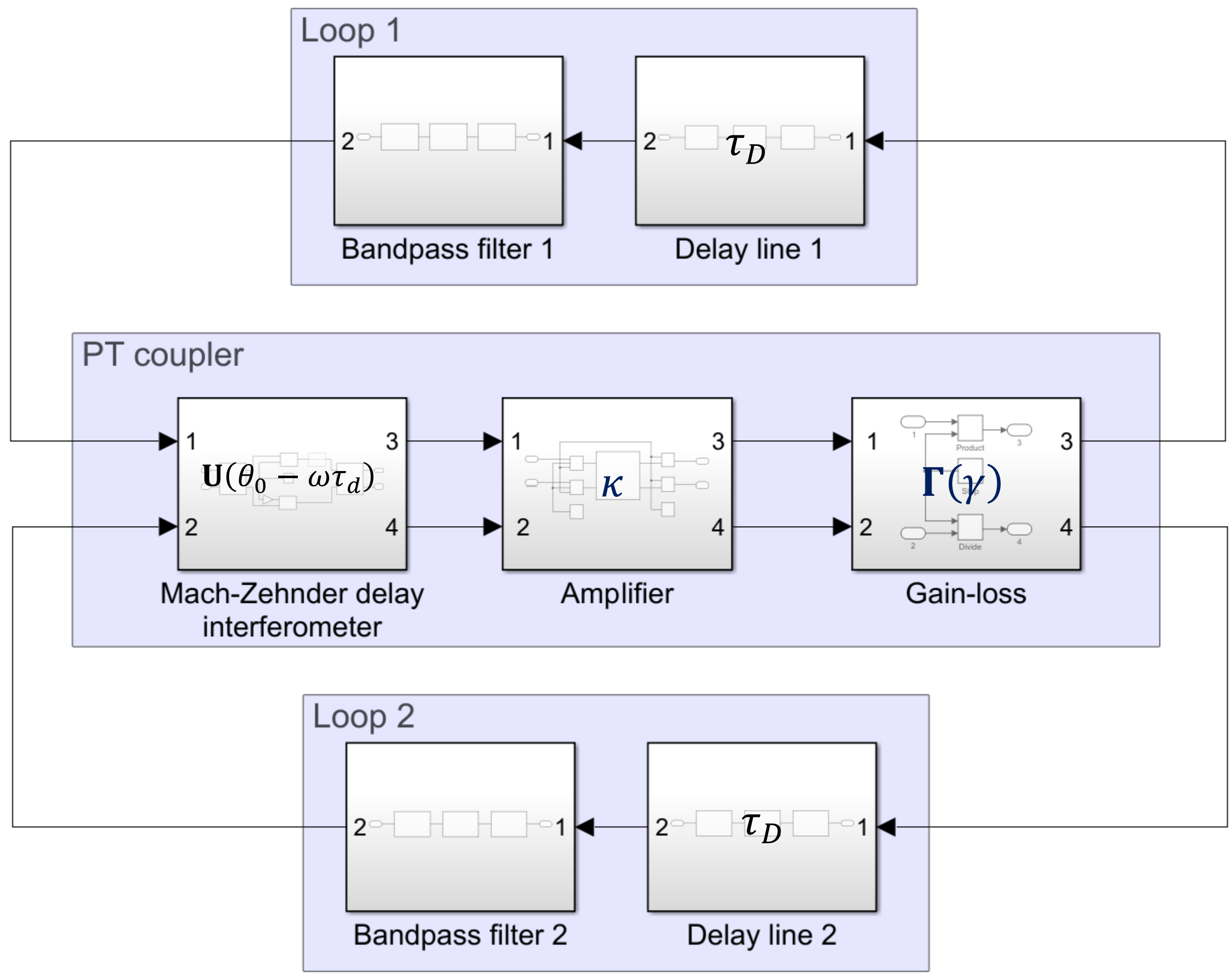

Figure 4 Block diagram of the complete gain-loss architecture.

# 3. Architectural interpretation

The two representations $\mathbf{K}_{\mathrm{I}}, \mathbf{K}_{\mathrm{II}}$ are *algebraically and physically equivalent* but have radically different *architectural* interpretation. The conjugation by $\mathbf{H}$ of the transmission matrix of two independent delay lines with distinct delays $\mathbf{D}_{\tau_+,\tau_-}$ illustrated in Figure 3 is key to comprehension of the equivalence of these distinct architectures.

## Gain-loss architecture

Equation 5 may be interpreted as describing two identical delay loops, represented by $\mathbf{D}_{\tau_D,\tau_D}(\omega)$, that include balanced gain-loss represented by $\mathbf{\Gamma}(\gamma)$, that are interconnected by a dispersive coupler represented by $\mathbf{U}(\theta_0 - \omega\tau_d)$. The frequency-dependent phase shift $\omega\tau_d$ introduces mode-dependent coupling, enabling frequency-selective $\mathcal{PT}$-symmetry transitions. The decomposition of $\mathbf{U}(\theta)$ in Equation 9 corresponds to the transmission matrix of a Mach-Zehnder interferometer (MZI) formed by two 50:50 180° hybrid couplers, represented by the Hadamard matrices $\mathbf{H}$, with a differential phase shift of $\theta$ within its arms. The non-dispersive component $\theta_0$ of the coupling is provided by a differential phase shift of $\theta_0$ within the two arms. The dispersive component $-\omega\tau_d$ is provided by a short delay imbalance in the arms of the interferometer (see Figure 3). Figure 4 illustrates schematically the complete gain-loss architecture.

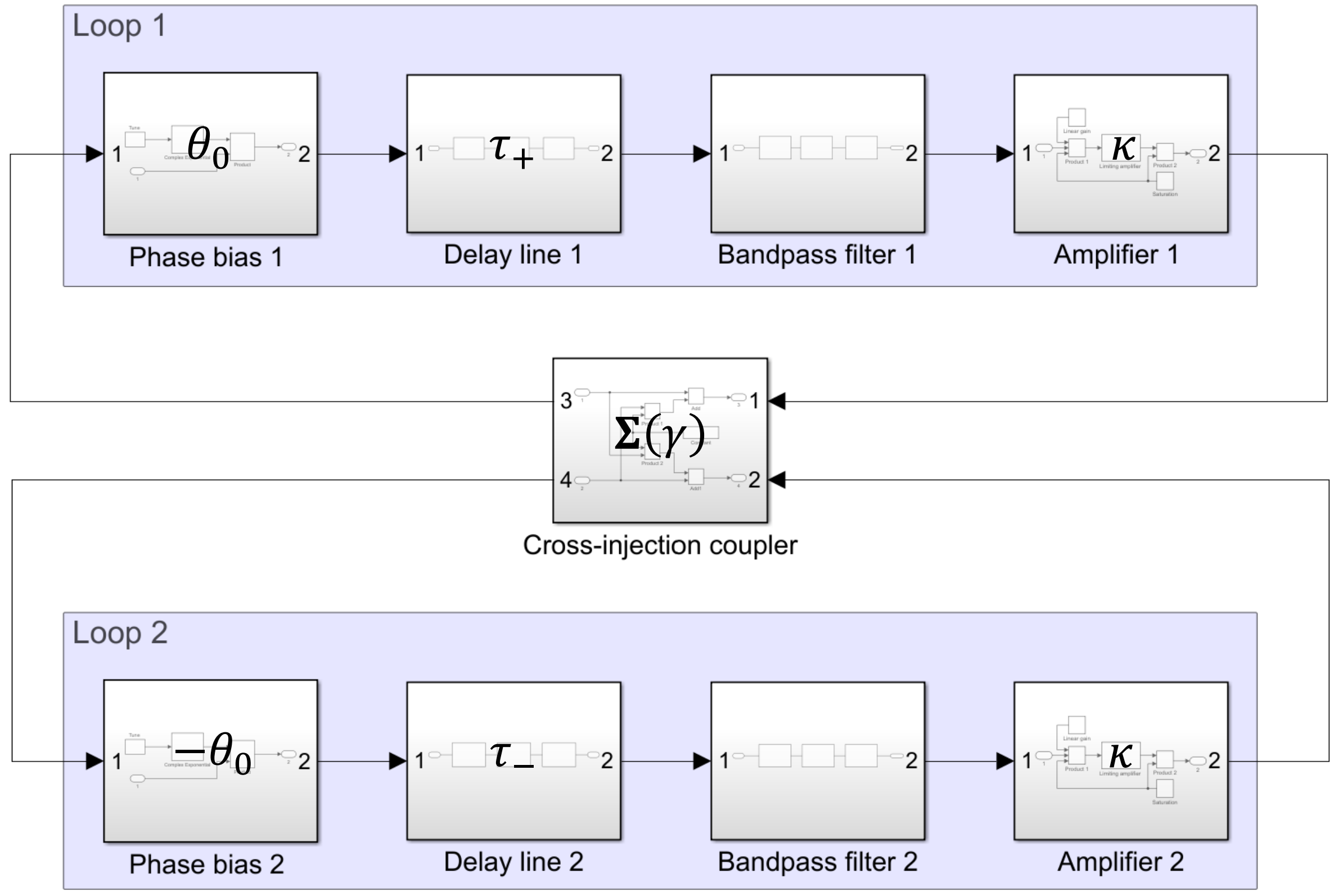

*Figure 5 Block diagram of the complete cross-injection architecture.*

## Cross-injection architecture

Equation 6 may be interpreted as describing two *lossless* loops with *distinct delay times*, represented by $\mathbf{D}_{\tau_+,\tau_-}(\omega)$, that include tuning phase shifts, with a differential phase shift of $\pm\theta_0$. The two loops are interconnected by a coupler, represented by the hyperbolic rotation matrix $\mathbf{\Sigma}$, which provides symmetric cross-injection with an injection ratio $\rho = \tanh(\gamma)$. Figure 5 illustrates schematically the complete cross-injection architecture.

The distinct free-spectral ranges of the loops give rise to a Vernier effect in which the relative alignment of resonances varies with frequency. In this picture, mode-selective $\mathcal{PT}$-symmetry transitions arise naturally from the periodic coincidence of resonances as illustrated in Figure 6.

## Summary

Table 1 compares the two architectures. Notably, the coupling between the two loops is determined by $\theta$ for the gain-loss architecture and by $\gamma$ for the cross-injection architecture.

| | | **gain / loss architecture** | **cross-injection architecture** |
|---|---|---|---|
| **loops** | | | |
| | **delay** | *matched* ($\tau_D$) | *distinct* $\tau_\pm = \tau_D \pm \tau_d$ |
| | **gain/loss** | *balanced* ($\pm\gamma$) | *none* |
| | **detuning** | *none* | *differential* ($\pm\theta$) |
| **coupling** | | *unitary* ($\mathbf{U}(\theta)$) | *hyperbolic* ($\mathbf{\Sigma}(\gamma)$) |
| **$\mathcal{PT}$- transition selectivity** | | *dispersion* $\theta(\omega)$ | *Vernier effect* |

*Table 1 Distinguishing features of the loss/gain and cross-injection architectures.*

For $\theta = 0$ the gain-loss architecture dissociates into two independent loops, with balanced gain/loss, identical delay ($\tau_D$) and zero detuning. For $\gamma = 0$ the cross-injection architecture dissociates into two independent lossless loops, with dispersive differential detuning $\theta(\omega)$. The magnitude of the (net) common gain $|\kappa|$ sustains oscillation and its phase $\arg(\kappa)$ provides (common) tuning. In practice, after the initial oscillation buildup where $|\kappa| > 1$, a gain control mechanism, typically amplifier saturation, stabilizes the magnitude of the oscillation so that $|\kappa| \longrightarrow 1$.

# 4. Vernier interpretation

The cross-injection architecture may be understood most naturally in terms of the Vernier effect. In the absence of cross-injection ($\rho = \tanh\gamma = 0$), the two oscillators possess distinct delay times $\tau_+$ and $\tau_-$, and therefore different free-spectral ranges. Their resonances form two frequency combs whose relative alignment varies periodically across the spectrum. At frequencies where neighbouring resonances coincide, the two oscillators interact most strongly. Conversely, where the resonance spacing is large, the interaction is weak.

The $\mathcal{PT}$-symmetric coupling acts on this Vernier structure rather than independently of it. As resonance coincidence improves, the modal frequency splitting decreases until a local PT-symmetry transition becomes possible. The broken $\mathcal{PT}$-symmetry intervals are therefore centred on the regions of strongest Vernier alignment. Mode selection emerges because those intervals preferentially amplify one member of the coupled resonance pair while suppressing its counterpart.

A previous work [13] introduced a $\mathcal{PT}$-symmetry order-parameter $\xi$ defined by:

$$\xi = \cosh(\gamma)\cos(\theta)$$

*Equation 13*

The exceptional points contour $|\xi| = 1$ delineates unbroken $\mathcal{PT}$-symmetry regimes $|\xi| < 1$ from broken $\mathcal{PT}$-symmetry regimes $|\xi| > 1$. The broken $\mathcal{PT}$-symmetry regions occur within intervals of $\theta$ centred on integer multiples of $\pi$. For small $\gamma$ these intervals are approximately:

$$\theta \in [-\gamma, \gamma] + p\pi \quad ; \quad p \in \mathbb{Z}$$

*Equation 14*

Near a frequency $\omega_0$ at which resonances of the two combs coincide ($\theta = 0$), one may write.

$$\theta(\omega) = -\omega\tau_d$$

*Equation 15*

Since the gain splitting vanishes at the exception points at the edge of the region, one can safely set the interval to span the averaged free-spectral range so that at the edges of the interval:

$$\omega\tau_D = \pm 2\pi$$

*Equation 16*

This leads to the requirement:

$$\gamma = \cosh^{-1}(1/\cos(2\pi\tau_d/\tau_D))$$

*Equation 17*

or approximately:

$$\gamma \sim 2\pi\tau_d/\tau_D \quad ; \quad \tau_d \ll \tau_D$$

*Equation 18*

For $\tau_D = 25\ \mu s, \tau_d = 0.5\ \mu s,$ Equation 17 and Equation 18 agree on $\gamma = 0.126$ to 3 significant figures.

The Vernier effect determines the frequency dependence of the $\mathcal{PT}$-symmetry transition. Resonance coincidence reduces the modal frequency splitting, allowing the system to enter the broken $\mathcal{PT}$ -symmetry regime within localised spectral intervals. Away from coincidence the resonance splitting increases, the $\mathcal{PT}$-transition disappears, and the coupled modes remain in the unbroken regime. Consequently, Vernier alignment acts as a selector that determines where along the spectrum $\mathcal{PT}$-symmetry breaking can occur. This interplay between the dispersive $\mathcal{PT}$-symmetry transitions and the Vernier alignment of the resonances of the uncoupled oscillators is illustrated schematically in *Figure 6*.

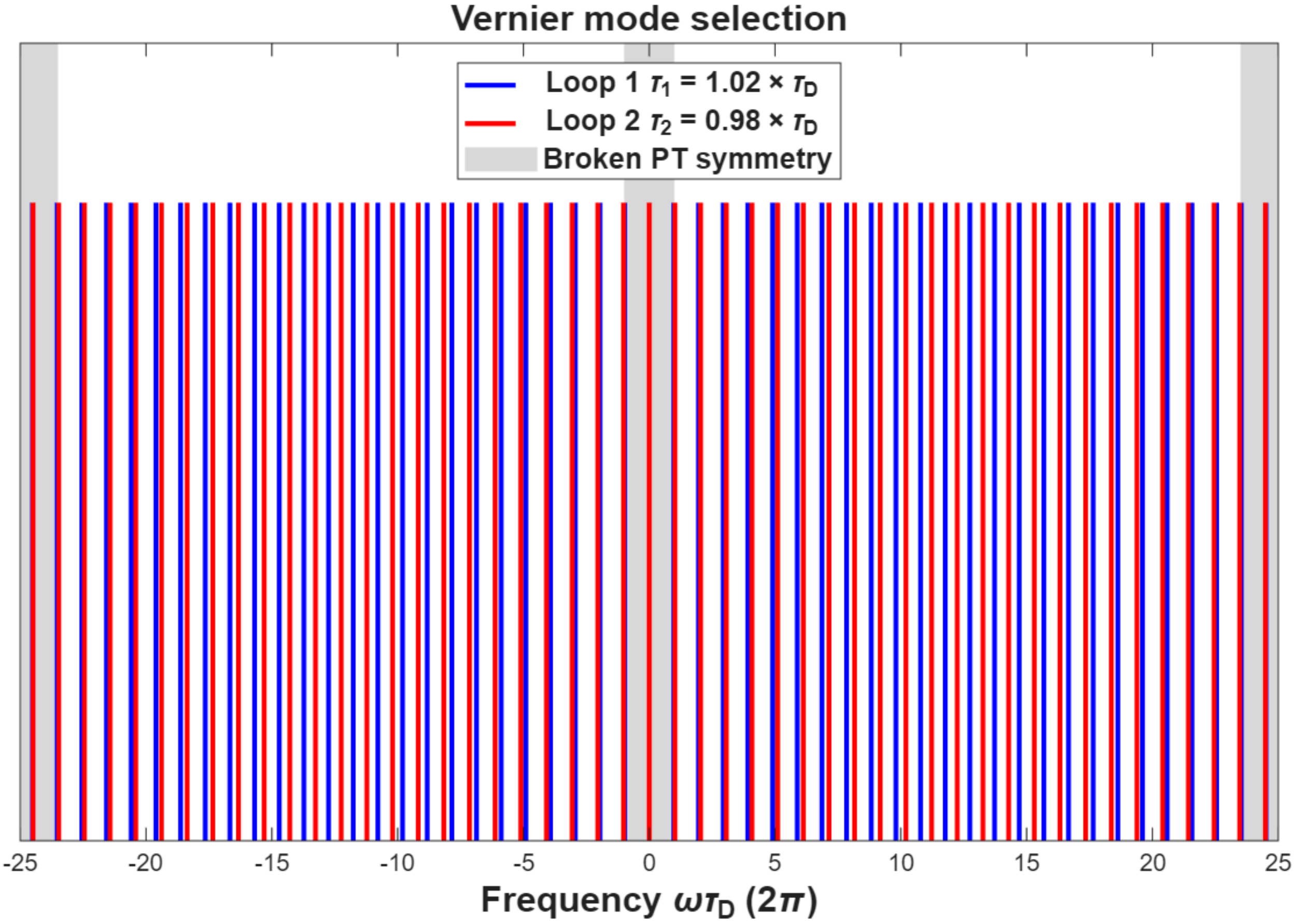

*Figure 6 Illustration of the interplay between dispersive $\mathcal{PT}$-symmetry transitions and the Vernier effect of the uncoupled $\rho = tanh(\gamma) = 0$ distinct resonant frequency combs of the component oscillators, for $\tau_+ = 1.02\tau_D$, $\tau_- = 0.98\tau_D$. The splitting between the resonances centered on the regular grid $\omega\tau_D = 2p\pi$, $p \in \mathbb{Z}$ varies periodically across the spectrum and is least in the greyed regions of broken $\mathcal{PT}$-symmetry where the resonances coalesce into degenerate pairs with reciprocal gain-loss ($\gamma = 0.126$). In the regions of unbroken $\mathcal{PT}$-symmetry the frequency splitting is too great for degenerate coalescence and the resonant pairs form doublets. The regions of closest Vernier alignment coincide with localised broken $\mathcal{PT}$-symmetry intervals and therefore act as preferred mode-selection regions for the nonlinear oscillator.*

Notably, in the saturation regime the two oscillators will injection lock provided the frequency splitting is less than the locking range region that is centred and contained within the broken $\mathcal{PT}$-symmetry region. Injection locking is a consequence of the nonlinearity introduced by gain saturation, while the $\mathcal{PT}$-symmetry is a consequence of a linear system with the extra degree of freedom provided by a two-dimensional oscillator state space. From the Vernier perspective the broken $\mathcal{PT}$-symmetry is an oscillator state-preparation process consistent with but distinct from classical injection locking.

# 5. Simulations

The objective of the simulations is not to re-establish the $\mathcal{PT}$-symmetric eigenvalue structure derived analytically in [13], but to demonstrate that the cross-injection architecture reproduces the predicted mode-selection behaviour in the nonlinear oscillation regime. Representative results are presented for the cross-injection architecture using a custom Simulink™ simulation platform. The Simulink simulation model is the same as shown in Figure 5 except that white noise initial condition and power-law phase fluctuation blocks were added to the delay blocks to model internal noise processes to generate realistic power and phase noise spectra. A Hadamard coupler is used to enable access to the dark and bright ports of the core oscillator (Figure 2(c)). Reference [9] provides detailed description of the inner workings of the principal subsystem blocks. The values of the principal parameters are listed in Table 2.

| | |
|---|---|
| $\tau_D = 24.9\ \mu s$ | *Common delay time* |
| $\tau_d = 0.5\ \mu s$ | *Differential delay time* |
| $\tau_R = 0.1\ \mu s$ | *Resonator on-resonance group delay* |
| $\tau_G = \tau_D + \tau_R = 25\ \mu s$ | *Common round trip group delay time* |
| $\gamma = 0.126$ | *Balanced gain/loss* |
| $\rho = \tanh(\gamma) = 0.125$ | *Injection ratio* |
| $FSR = 1/\tau_G = 40\ kHz$ | *Regular frequency grid interval* |
| $\Delta f = 1/\pi\tau_R = 3.18\ MHz$ | *Resonator bandwidth* |

*Table 2 Principal simulation parameter values*

The analysis developed in Section 2 is linear and describes the oscillation build-up process. Numerical simulations and experimental realizations require the oscillation amplitude to stabilize through gain saturation. In the simulations presented here, saturation is modelled using conventional amplifier saturation characteristics [9]. For the cross-injection architecture, satisfactory operation is obtained using a pair of nominally identical independent amplifiers, since the final oscillatory state occupies both loops with comparable amplitudes.

In contrast, architectures based on explicit gain-loss loops are considerably more sensitive to the details of the saturation process because unequal modal amplitudes can induce unequal saturated gains. Maintaining balanced gain and loss in the nonlinear regime may therefore require a common saturation mechanism acting on the combined oscillator state.

The influence of saturation mechanisms on the preservation of $\mathcal{PT}$-symmetry in the nonlinear regime is beyond the scope of the present work. A detailed analysis of these effects is deferred to future work.

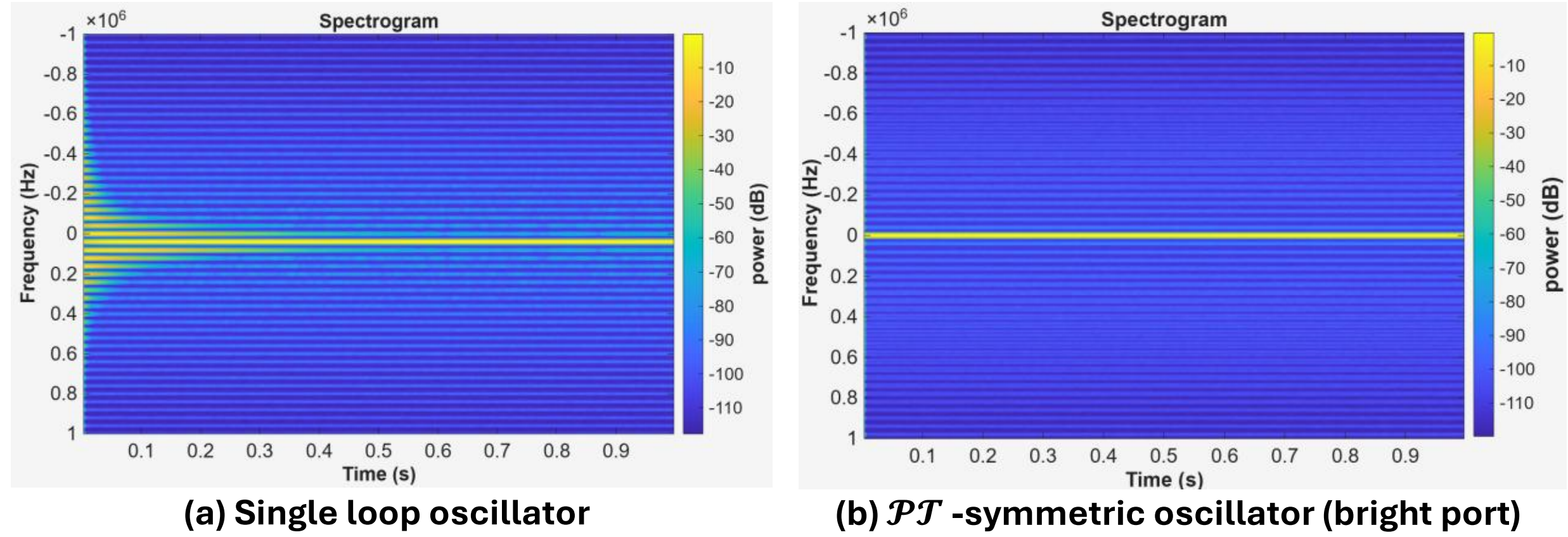

**(a) Single loop oscillator** **(b) $\mathcal{PT}$ -symmetric oscillator (bright port)**

*Figure 7. Comparison of the evolution towards a single mode oscillating state of the $\mathcal{PT}$ — symmetric oscillator (cross-injection architecture) and a single loop oscillator with the same phase fluctuations, initial conditions parameters and similar round-trip delay ($25\ \mu s$). The single loop oscillator requires $\sim 500\ ms$ for its sidemodes to decay to a noise-driven residual level, while the $\sim 2ms$ initial transient of the $\mathcal{PT}$ — symmetric oscillator is too brief to be visible in the RF spectrogram.*

Figure 7 presents two spectrograms (short time Fourier transforms) that contrast the evolution of the power spectrum of the oscillations of a single loop oscillator and the $\mathcal{PT}$-symmetric oscillator. The sidemodes of the initial oscillation of the single-loop oscillator decay slowly; a good proportion of a second elapses before the sidemodes reach a residual level driven by noise. The regular frequency interval between modes is clearly visible. In contrast, the initial transient of the $\mathcal{PT}$-symmetric oscillator is too brief to be resolved by the spectrogram which appears visually uncorrupted by noise.

The characteristic doublet structure of the $\mathcal{PT}$-symmetric phase and the slow periodic variation of the doublet splitting-frequency are just visible in Figure 7. These characteristic distinguishing features of the single mode and oscillator and the $\mathcal{PT}$-symmetric oscillator are clearer in the power spectrum of the respective oscillations shown in Figure 8.

It can also be observed in the case of the single-loop oscillator that the mode-competition process has selected a mode displaced by one free-spectral range from the intended main mode at zero offset frequency, while the $\mathcal{PT}$-symmetric oscillator consistently selects the mode associated with the strongest Vernier alignment and corresponding local broken $\mathcal{PT}$-symmetry region. Competing modes remain within the unbroken $\mathcal{PT}$-symmetry regime and therefore retain finite frequency splitting, reducing their ability to dominate the mode-competition process.

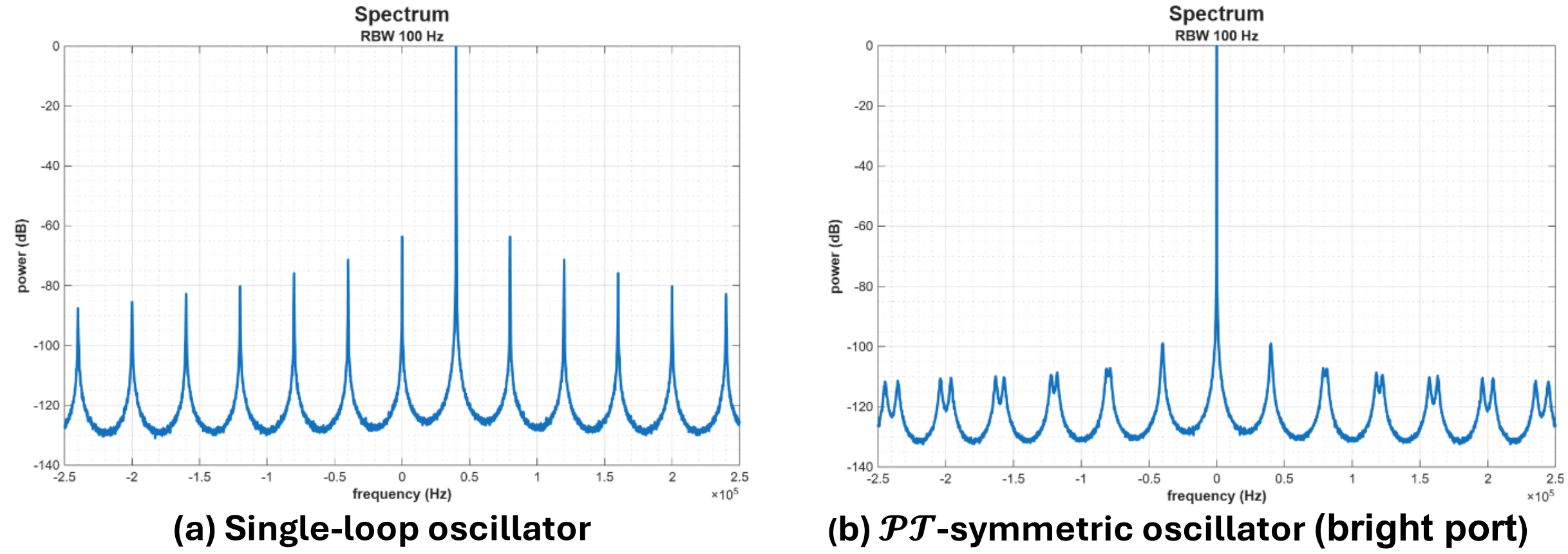

**(a) Single-loop oscillator** **(b) $\mathcal{PT}$-symmetric oscillator (bright port)**

*Figure 8. Comparison of the asymptotic power spectral density of the $\mathcal{PT}$-symmetric oscillator (cross-injection architecture) and a single loop oscillator with the same phase fluctuations, initial conditions parameters similar round-trip delay (25 $\mu s$). The unbroken $\mathcal{PT}$-symmetry doublet resonances are clearly visible in the power spectral density of the $\mathcal{PT}$-symmetric oscillator. In the case of the single-loop oscillator the mode competition process starting from noise has resulted in an oscillating mode displaced from the resonator passband centre frequency by one FSR while the mode-selective $\mathcal{PT}$-symmetry phase transition results in an oscillating mode at the passband centre.*

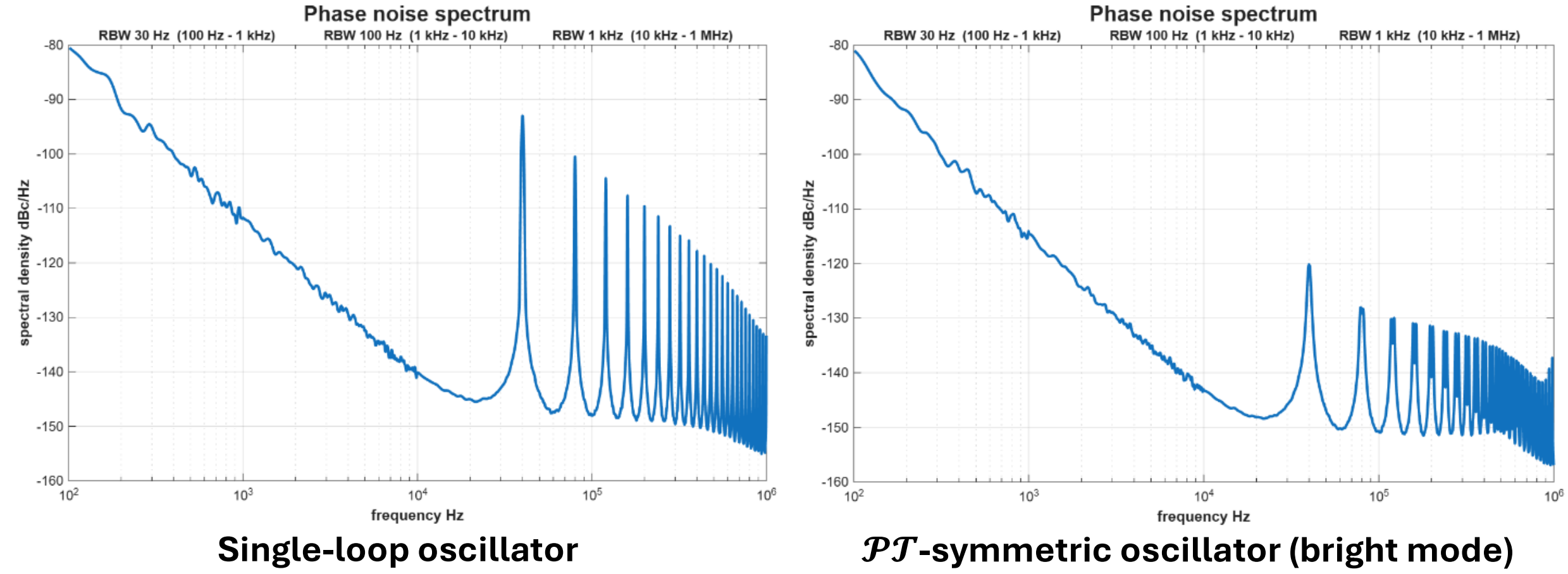

**Single-loop oscillator** **$\mathcal{PT}$-symmetric oscillator (bright mode)**

*Figure 9. Comparison of the phase noise spectral density of the $\mathcal{PT}$-symmetric oscillator (cross-injection architecture) and a single loop oscillator with the same phase fluctuation, initial condition and similar round-trip delay (25 $\mu s$). The $\mathcal{PT}$-symmetric oscillator spurious sidemodes are significantly suppressed relative to the single-loop oscillator.*

Figure 9 presents a comparison of the phase noise spectral density of the data generated by the single loop and $\mathcal{PT}$-symmetric oscillator simulation models. It is observed that the $\mathcal{PT}$-symmetric oscillator has a 3 dB lower phase noise than the single loop oscillator at offset frequencies below the first sidemode resonance, but the sidemode level oof the $\mathcal{PT}$-symmetric oscillator is substantially lower (by 27 dB for the first sidemode) than the sidemode level of the single-loop oscillator.

# 6. Conclusions

A generalized formulation of $\mathcal{PT}$-symmetric time-delay oscillators has been developed in which the round-trip operator is permitted to be dispersive. Unlike previous formulations based on matched delay loops and frequency-independent coupling, the generalized framework supports localized, mode-selective $\mathcal{PT}$-symmetry transitions and does not require matched delays.

Two mathematically equivalent representations of the resulting $\mathcal{PT}$-symmetric oscillator have been derived. One corresponds to coupled gain-loss loops interconnected by a dispersive coupler. The other corresponds to a pair of equal-gain oscillators with unequal delay times coupled by symmetric cross-injection. The equivalence of these representations establishes a direct relationship between $\mathcal{PT}$-symmetric dynamics and practical dual-loop oscillator architectures.

The cross-injection representation provides an intuitive interpretation of mode-selective $\mathcal{PT}$- symmetry in terms of the Vernier effect. In this picture, localized symmetry transitions arise when resonances of the uncoupled oscillators approach coincidence, producing frequency-selective gain-loss splitting and robust mode selection. $\mathcal{PT}$-symmetry therefore appears not merely as an abstract symmetry principle, but as the mechanism by which Vernier alignment is converted into robust frequency-selective mode selection. Mode-selective $\mathcal{PT}$-symmetry is the manifestation of Vernier-assisted cross-injection between oscillators with unequal delays.

Numerical simulations confirm the theoretical predictions and demonstrate rapid convergence to a single-mode oscillation state, substantial sidemode suppression, and preservation of low phase-noise performance. These results provide a practical framework for the design of $\mathcal{PT}$-symmetric microwave-photonic oscillators and suggest new connections between non-Hermitian dynamics, cross-injection architectures, and Vernier-based frequency selection.

# Appendix I

## Mathematical framework

A pair of time delay oscillators coupled by linear time invariant network represented by the operator $\boldsymbol{\mathcal{K}}$ is described in the time domain by:

$$\boldsymbol{u} = \kappa \boldsymbol{\mathcal{K}} \boldsymbol{u}$$

*Equation 19*

where $\boldsymbol{u} \in \mathbb{C}^2$ is the oscillator state vector and $\kappa \in \mathbb{C}$ represents the common roundtrip gain and phase. $\boldsymbol{\mathcal{K}}$ is characterised in the frequency domain by its transmission matrix $\mathbf{K}(\omega)$ which now encodes time delay through its frequency ($\omega \in \mathbb{R}$) dependence.

Following [13], set:

$$\mathbf{K} = \mathbf{HWH} \quad ; \quad \mathbf{H} = \frac{1}{\sqrt{2}}(\sigma_1 + \sigma_3)$$

*Equation 20*

then $\mathbf{K}$ is $\mathcal{PT}$-symmetric if:

$$\mathbf{KJK}^\dagger = \mathbf{J} \quad \Leftrightarrow \quad \mathbf{W}\boldsymbol{\eta}\,\mathbf{W}^\dagger = \boldsymbol{\eta}$$

*Equation 21*

where $\sigma_1, \sigma_3$ are the first and third of the three Pauli matrices:

$$\sigma_1 = \begin{bmatrix} 0 & 1 \\ 1 & 0 \end{bmatrix} \quad ; \quad \sigma_2 = \begin{bmatrix} 0 & -i \\ i & 0 \end{bmatrix} \quad ; \quad \sigma_3 = \begin{bmatrix} 1 & 0 \\ 0 & -1 \end{bmatrix}$$

*Equation 22*

$\mathbf{J} = \sigma_1$ is the exchange matrix in two dimensions, $\boldsymbol{\eta} = \sigma_3$ is the indefinite metric in signature form $(1, -1)$, and $\mathbf{W}$ is an element of the pseudo-unitary group of transformations $U(1{,}1)$ that preserve the indefinite metric $\boldsymbol{\eta}$.

## Decomposition

$\mathbf{W}$ admits the Cartan (KAK) decomposition of $U(1{,}1)$:

$$\mathbf{W} = \mathbf{V}_1\boldsymbol{\Sigma}\mathbf{V}_2$$

*Equation 23*

where $\mathbf{V}_{1,2}$ take the form:

$$\mathbf{V} = \begin{bmatrix} \exp(i\varphi_+) & 0 \\ 0 & \exp(i\varphi_-) \end{bmatrix}$$

*Equation 24*

and $\boldsymbol{\Sigma}$ takes the form:

$$\boldsymbol{\Sigma} = \begin{bmatrix} \cosh(\gamma) & \sinh(\gamma) \\ \sinh(\gamma) & \cosh(\gamma) \end{bmatrix}$$

*Equation 25*

where $\mathrm{V}_+, \mathrm{V}_-$ carry the compact degrees of freedom and $\boldsymbol{\Sigma}$ carries the non-compact degree of freedom $(\gamma)$.

Any decomposition of $\mathbf{W}$ into a product of factors preserves the metric $\boldsymbol{\eta}$ provided each factor separately preserves the same metric.

1) Although $\mathbf{K} = \mathbf{MWM^{-1}}$ is formally a similarity transformation, the matrices $\mathbf{M}$ and $\mathbf{M^{-1}}$ correspond to ideal feedback connections that cancel in the round-trip operator. Consequently, the transformed systems $\mathbf{K} = \mathbf{MWM^{-1}}$ and $\mathbf{K} = \mathbf{W}$ are physically as well as algebraically equivalent.

2) The hyperbolic rotation $\boldsymbol{\Sigma}$ is equal to the conjugation by $\mathbf{H}$ of a diagonal gain-loss matrix $\boldsymbol{\Gamma}$:

$$\mathbf{\Sigma} = \mathbf{H\Gamma H} \quad ; \quad \mathbf{\Gamma}(\gamma) = \exp(\gamma\sigma_3)$$

*Equation 26*

3) A pair of delay lines, with *distinct* delay times $\tau_+$, $\tau_-$, is described by a diagonal transmission matrix $\mathbf{D}$ defined by:

$$\mathbf{D}_{\tau_+,\tau_-}(\omega) = \begin{bmatrix} \exp(-i\omega\tau_+) & 0 \\ 0 & \exp(-i\omega\tau_-) \end{bmatrix} = \exp(i\omega\tau_d\sigma_3)\, \mathbf{D}_{\tau_D,\tau_D}(\omega)$$

*Equation 27*

that trivially preserves the metric $\boldsymbol{\eta}$.

Where:

$$\tau_\pm = \tau_D \pm \tau_d$$

*Equation 28*

4) The conjugation by $\mathbf{H}$ of a diagonal phase shift matrix $\mathbf{\Theta}$ defined by:

$$\mathbf{\Theta}(\theta) = \exp(i\theta\sigma_3)$$

*Equation 29*

is a bisymmetric special unitary matrix:

$$\mathbf{U}(\theta) = \mathbf{H\Theta}(\theta)\mathbf{H} \equiv \begin{bmatrix} \cos(\theta) & i\sin(\theta) \\ i\sin(\theta) & \cos(\theta) \end{bmatrix}$$

*Equation 30*

## Two equivalent representations

Applying these rules and definitions, one may construct two representations $\mathbf{K}_\mathbf{I}$, $\mathbf{K}_\mathbf{II}$ of the equivalence class of $\mathcal{PT}$-symmetric roundtrip transmission matrices $\mathbf{K}$.

$$\mathbf{K}_\mathbf{I}(\omega) = \mathbf{\Gamma}(\gamma)\mathbf{U}(\theta_0 - \omega\tau_d)\mathbf{D}_{\tau_D,\tau_D}(\omega)$$

*Equation 31*

and:

$$\mathbf{K}_\mathbf{II}(\omega) = \mathbf{\Sigma}(\gamma)\mathbf{\Theta}(\theta_0)\mathbf{D}_{\tau_+,\tau_-}(\omega)$$

*Equation 32*

Where $\theta_0$ in Equation 31 may be interpreted as the non-dispersive component of the coupling and in Equation 32 as the differential component of the tuning phase shifts.

Each factor $\mathbf{\Gamma}$, $\mathbf{U}, \mathbf{D}_{\tau_D,\tau_D}$ in Equation 31 separately satisfies the $\mathcal{PT}$-symmetry condition $\mathbf{KJK}^\dagger = \mathbf{J}$ and consequently their composition $\mathbf{K}_\mathbf{I}$ is $\mathcal{PT}$-symmetric. Each factor $\mathbf{\Sigma}$, $\mathbf{\Theta}$, $\mathbf{D}_{\tau_+,\tau_-}$ in Equation 32 separately satisfies $\mathbf{W}\boldsymbol{\eta}\,\mathbf{W}^\dagger = \boldsymbol{\eta}$ and consequently their composition $\mathbf{K}_\mathbf{II}$ preserves the metric. It follows that $\mathbf{K}_\mathbf{I}$ is related to $\mathbf{K}_\mathbf{II}$ by similarity:

$$\mathbf{K}_\mathbf{I} = \mathbf{HK}_\mathbf{II}\mathbf{H}$$

*Equation 33*

Consistent with 1) in the proceeding, the core oscillator is preserved by similarity and only the external observation of the oscillator state-vector is transformed (see Figure 2).